\documentclass[12pt]{iopart}
\usepackage{epsfig}

\def\dddot#1{\mathinner{\buildrel\vbox{\kern5pt\hbox{...}}\over{#1}}}

\def\bs{\begin{subequations}}
\def\es{\end{subequations}}
\def\be{\begin{equation}}
\def\ee{\end{equation}}
\def\bq{\begin{eqnarray}}
\def\eq{\end{eqnarray}}
\def\beq{\begin{eqnarray*}}
\def\eeq{\end{eqnarray*}}
\def\ba{\begin{eqnarray}}
\def\ea{\end{eqnarray}}

\begin{document}

\title{Noether and Lie symmetries for charged perfect fluids}

\author{ M C Kweyama\dag\
 \footnote[2]{ Permanent address: Department of Mathematical Sciences,
 Mangosuthu University
of Technology,  P. O. Box 12363, Jacobs  4026, South Africa},
  K S Govinder\dag  and S D Maharaj\dag}
\address{\dag\ Astrophysics and Cosmology Research Unit,
 School of Mathematical Sciences,
 University of KwaZulu-Natal, Private Bag X54001,
 Durban 4000, South Africa \\ Email: maharaj@ukzn.ac.za
 }

\begin{abstract}
We study the underlying nonlinear partial differential equation that governs the behaviour of spherically
symmetric charged fluids in general relativity.  We investigate the conditions for the equation to admit a first integral or be reduced to quadratures
using symmetry methods for differential equations.  A general Noether first integral is found. We also undertake a comprehensive group analysis of the underlying equation
using Lie point symmetries. The existence of a Lie symmetry is subject to solving an integro-differential
equation in general;  we investigate the conditions under which it can be reduced to quadratures.  Earlier results for uncharged fluids and particular 
first integrals for charged matter
are regained as special cases of  our treatment.
\end{abstract}


\maketitle

\section{Introduction}
The Einstein-Maxwell system of equations plays a central role in relativistic astrophysics when describing 
spherically symmetric gravitational fields in static manifolds.  In these situations we are modelling  charged
 compact objects with strong gravitational fields such as dense neutron stars.  Some recent investigations, 
 including the models of Ivanov \cite{Ivanov}, Sharma \textit{et al} \cite{Sharma} and Thirukkanesh and 
 Maharaj \cite{Thirukkanesh}, indicate that the electromagnetic field significantly affects physical quantities in 
 relativistic stellar systems: equations of state, redshifts, luminosities, stability and maximum masses of compact 
 relativistic stars.  The presence of electric charge is a necessary ingredient in the structure and gravitational evolution
  of stars composed of quark matter as illustrated in the treatments of Mak and Harko \cite{Mak} and Komathiraj and 
  Maharaj \cite{Komathiraj}.  These applications emphasise the need for exact solutions of the Einstein-Maxwell system 
  in modelling the formation and evolution of charged astrophysical objects.  Lasky and Lun \cite{Lasky, Las}, amongst
   others, studied the role of electromagnetic fields in gravitational collapse, formation of black holes and the existence 
   of naked singularities.  Electric fields cannot be ignored in spherical gravitational collapse with phases of intense 
   dynamical activity and particle interaction as shown in the treatments of Di Prico \textit{et al} \cite{Di} and 
   Herrera \textit{et al} \cite{Herrera}.  Maxwell's equations also play an important role in cosmological models in 
   higher dimensions, brane world models and wormhole configurations.  In this context see the treatment of De Felice and Ringeval \cite{De}.

When solving the Einstein field equations with neutral matter distributions, we often make the assumption that the spacetime is 
shear-free and spherically symmetric.  Kustaanheimo and Qvist \cite{Kustaanheimo} were the first to present a general class of solutions.  
The generalisation to include the presence of the electromagnetic field may be easily achieved.  The field equations are reducible to 
a single partial differential equation.  A review of known charged solutions, with a Friedmann limit, is given by 
Krasinski \cite{Krasinski}.  Srivastava \cite{Srivastava} and Sussmann \cite{Suss88a, Suss88b} undertook a detailed study 
of the mathematical and physical features of the Einstein-Maxwell system in spherical symmetry.  Wafo Soh and Mahomed \cite{Wafo1}
 used symmetry methods to systematically study the underlying partial differential equation.  
 They showed that all previously known solutions can be related to a Noether point symmetry

The main aim of this paper is to study the integrability properties of the underlying partial differential equation for the Einstein-Maxwell 
system using symmetry methods.  Both Noether and Lie point symmetries of the governing equation are considered.  
Noether symmetries have the interesting property of being associated with physically relevant conservation laws in a direct manner
via the well-known 
Noether theorem.  The Lie symmetries are more general, providing a larger set of symmetry generators in general, 
but do not guarantee integrability and reduction to quadrature in a straight forward manner.  In section 2, we reduce the Einstein-Maxwell
 field equations to a single nonlinear second order partial differential equation that governs the behaviour of charged fluids.  
 This is achieved by utilising the generalised transformation due to Faulkes \cite{Faulkes}.  The resulting partial differential equation
  can be treated as an ordinary differential equation as in the case of uncharged fluids.  In section 3, we present a first integral of the governing
   equation obtained earlier by generalising the technique of Srivastava \cite{Srivastava} first used for uncharged fluids.  This first integral is subject to two
    integrability conditions expressed as nonlinear integral equations.  We transform the integral conditions into a fourth order differential 
    equation. In section 4, we analyse the governing equation for Noether symmetries via its Lagrangian and this analysis yields a general Noether first integral for this
     equation.  We then establish the relationship between the Noether first integral and the first integral (which
     had been obtained earlier using an \textit{ad hoc} 
     approach).  In section 5, we undertake a comprehensive Lie symmetry analysis of the governing equation to investigate the conditions 
     under which it can be reduced to quadratures.  We show how the Noether results are a subset of the Lie analysis results. Lastly, in section 6, we discuss the results obtained and relate some invariant
     solutions to known results.

\section{Field equations}
We analyse the shear-free motion of a fluid distribution in the presence of an electric field.  
It is possible to choose coordinates $x^{i}=(t,r,\theta,\phi)$ such that the line element can be written in the  form
\begin{equation}
ds^{2}=-e^{2\nu(t,r)}dt^{2}+e^{2\lambda(t,r)}\left[dr^{2}+r^{2}\left(d\theta^{2}
+\sin^{2}\theta d\phi^{2}\right)\right]\label{eqna01}
\end{equation}
which is simultaneously comoving and isotropic.  Since an electromagnetic field is present, the Einstein
 field equations are supplemented with Maxwell's equations to describe a self-gravitating charged fluid. 
  The Einstein field equations for a charged perfect fluid can be written as the system
  \numparts
\begin{eqnarray}
\fl \rho=3\frac{\lambda_{t}^{2}}{e^{2\nu}}-\frac{1}{e^{2\lambda}}\left(2\lambda_{rr}
+\lambda_{r}^{2}+\frac{4\lambda_{r}}{r}\right)-\frac{E^{2}}{r^{4}e^{4\lambda}} \label{eqn01a}\\
\fl p=\frac{1}{e^{2\nu}}\left(-3\lambda_{t}^{2}-2\lambda_{tt}+2\nu_{t}\lambda_{t}\right)
+\frac{1}{e^{2\lambda}}\left(\lambda_{r}^{2}+2\nu_{r}\lambda_{r}+\frac{2\nu_{r}}{r}+\frac{2\lambda_{r}}{r}\right)+\frac{E^{2}}{r^{4}e^{4\lambda}} \label{eqn01b}\\
\fl p=\frac{1}{e^{2\nu}}\left(-3\lambda_{t}^{2}-2\lambda_{tt}+2\nu_{t}\lambda_{t}\right)
+\frac{1}{e^{2\lambda}}\left(\nu_{rr}+\nu_{r}^{2}+\frac{\nu_{r}}{r}+\frac{\lambda_{r}}{r}+\lambda_{rr}\right)-\frac{E^{2}}{r^{4}e^{4\lambda}} \label{eqn01c}\\
\fl 0=\nu_{r}\lambda_{t}-(\lambda_{t})_{r} \label{eqn01d}
\end{eqnarray}
\endnumparts
Maxwell's equations imply that
\begin{eqnarray}
E&=&r^{2}e^{\lambda-\nu} \Phi_{r}, \qquad E_{r} =\sigma r^{2}e^{3\lambda}\label{eqn02}
\end{eqnarray}
In the above $\rho$ is the energy density, $p$ is the isotropic pressure, $\sigma$ is the proper charge density 
and $E$ is the total charge contained within the sphere of radius $r$ centered around the origin.  The matter 
variables are measured relative to the four-velocity $\mu^{a}=\left(e^{-\nu},0,0,0\right)$.  Subscripts refer to partial 
derivatives.  Observe that $\Phi_{r}=F_{10}\left(=-F_{01}\right)$ is the only nonzero component of the Faraday 
electromagnetic field tensor $F_{ab}=\phi_{b;a}-\phi_{a;b}$ where we have chosen $\phi_{a}=\left(\Phi(t,r),0,0,0\right)$ 
as the simplest choice that allows for a nonvanishing electric field.  The coupled equations (\ref{eqn01a})-(\ref{eqn01d}), (\ref{eqn02})
comprise the Einstein-Maxwell system for the metric (\ref{eqna01}) in terms of the variables $\phi, p, E, \sigma, \nu, \lambda$.

The Einstein-Maxwell field equations can be reduced to a simpler system as shown by Srivastava \cite{Sriv}.
After considerable simplification the Einstein field equations (\ref{eqn01a})-(\ref{eqn01d}) can be written in the equivalent form
\numparts
\begin{eqnarray}
\rho=3e^{2h}-e^{-2\lambda}\left(2\lambda_{rr}+\lambda_{r}^{2}+\frac{4\lambda_{r}}{r}\right)-\frac{E^{2}}{r^{4}e^{4\lambda}}  \label{eqn09a}\\
p=\frac{1}{\lambda_{t}e^{3\lambda}}\left[e^{\lambda}\left(\lambda_{r}^{2}+\frac{2\lambda_{r}}{r}\right)-e^{3\lambda+2h(t)}-\frac{E^{2}e^{-\lambda}}{r^{4}}\right]_{t} \label{eqn09b}\\
e^{\nu}=\lambda_{t}e^{-h} \label{eqn09c}\\
e^{\lambda}\left(\lambda_{rr}-\lambda_{r}^{2}-\frac{\lambda_{r}}{r}\right)=-\tilde{F}-\frac{2E^{2}e^{-\lambda}}{r^{4}} \label{eqn09d}
\end{eqnarray}
\endnumparts
for a  charged relativistic fluid, where $h=h(t)$ and $\tilde{F}=\tilde{F}(r)$ are arbitrary constants of 
integration.  We have the condition $\lambda_t \neq 0$ from (\ref{eqn09b})
so that the model must be time dependent, and the spacetime cannot become static.
Observe that (\ref{eqn09d}) is essentially the generalised condition of pressure isotropy 
which includes a term corresponding to the electric field.  To find an exact solution to the system (\ref{eqn09a})-(\ref{eqn09d}) 
we need to explicitly integrate (\ref{eqn09d}) to determine $\lambda$, and $\sigma$ follows from (\ref{eqn02}).

To continue we need to express (\ref{eqn09d}) in a simpler form.  We use a transformation first 
utilised by Faulkes \cite{Faulkes} for neutral fluids.  We let
\numparts
\begin{eqnarray}
x=r^{2} \label{eqn010a}\\
y=e^{-\lambda}\label{eqn010b}\\
f(x)=\frac{\tilde{F}(r)}{4r^{2}} \label{eqn010c}\\
g(x)=\frac{E^{2}}{2r^{6}} \label{eqn010d}
\end{eqnarray}
\endnumparts
With the help of (\ref{eqn010a})-(\ref{eqn010d}) we can write (\ref{eqn09d}) in a more compact form
\begin{equation}
y''=f(x)y^{2}+g(x)y^{3} \label{eqn013}
\end{equation}
where primes denote differentiation with respect to the variable $x$.  Equation (\ref{eqn013}) is the fundamental nonlinear 
partial differential equation which determines the behaviour of the self-gravitating charged fluid in general relativity. However, we can treat it as an ordinary differential equation as only derivatives with respect to $x$ appear. 
 If $g=0$ then we regain a neutral fluid which has been studied by Maharaj {\it et al} \cite{Maharaj} and Wafo Soh 
 and Mahomed \cite{Wafo2}, amongst others.

\section{A charged first integral}
It is possible to find a first integral of (\ref{eqn013}) without choosing explicit forms of the 
functions $f(x)$ and $g(x)$.  We use integration by parts, an approach adopted by 
Maharaj {\it et al} \cite{Maharaj} in investigating the integrability properties of the field
 equation $y''=f(x)y^{2}$ of a neutral spherically symmetric shear-free fluid.  
 We can integrate (\ref{eqn013}) by parts to obtain
\begin{eqnarray}
y'&=&f_{I}y^{2}+g_{I}y^{3}-2f_{II}y y'+2f_{III}y'^{2}+2\left(ff_{II}\right)_{I}y^{3}+2\left(gf_{II}\right)_{I}y^{4}\nonumber\\
& &-K_{0}y^{3}-K_{1}y^{4}+\tau_{0}(t) \label{eqna10}
\end{eqnarray}
For convenience we have used the notation
\begin{eqnarray}
\int f(x)dx&=&f_{I} \qquad \int g(x)dx=g_{I} \label{eqna12}
\end{eqnarray}
The result (\ref{eqna10}) is subject to the integrability conditions
\numparts
\begin{eqnarray}
K_{0}=\frac{4}{3}ff_{III}+2\left(ff_{II}\right)_{I}+g_{I} \label{eqna11a}\\
K_{1}=gf_{III}+2\left(gf_{II}\right)_{I} \label{eqna11b}
\end{eqnarray}
\endnumparts
where $K_{0}$ and $K_{1}$ are constants and $\tau_{0}(t)$ is an arbitrary function of integration.

The integral equations (\ref{eqna11a})-(\ref{eqna11b}) may be simplified.  On setting
\begin{equation}
f_{III}=a \label{eqna13}
\end{equation}
we can rewrite (\ref{eqna11b}) as
\[ga+2\left(ga'\right)_{I}=K_{1}\]
which has the solution
\begin{equation}
g=g_{0}a^{-3} \label{eqna17}
\end{equation}
where $a=a(x)$ and $g_{0}$ is an arbitrary constant.  Equation (\ref{eqna11a}) now becomes
\[\frac{4}{3}a'''a+2\left(a'''a'\right)_{I}+\left(g_{0} a^{-3}\right)_{I}=K_{0}\]
or, if we want a purely differential equation,
\begin{equation}
aa^{(iv)}+\frac{5}{2}a'a'''=-\frac{3}{4}g_{0}a^{-3} \label{eqna20}
\end{equation}
We have therefore established that the integrability conditions (\ref{eqna11a})-(\ref{eqna11b}) can be transformed into the fourth order ordinary differential
 equation (\ref{eqna20}).  Using (\ref{eqna11a})-(\ref{eqna11b}), (\ref{eqna13}) and (\ref{eqna17}) we rewrite the first integral (\ref{eqna10}) of (\ref{eqn013}) as
\begin{equation}
\tau_{0}(t)=y'-a''y^{2}+2a'yy'-2ay'^{2}+\frac{4}{3}aa'''y^{3}+g_{0}a^{-2}y^{4} \label{eqna18},
\end{equation}
subject to condition (\ref{eqna20}).
  A solution of (\ref{eqna20}) will give $a$, and then $f$ and $g$ will be found from (\ref{eqna13}) and (\ref{eqna17}) respectively.  
  A full analysis of this case can be found in Kweyama {\it et al} \cite{Kweyama}, and we will not repeat those results here.

\section{Noether symmetries and integration}
Given that Noether's theorem generates first integrals in a direct manner, 
we now investigate (\ref{eqn013}) for Noether symmetries 
in order to generate first integrals. Note that, if a second order ordinary differential equation
\begin{equation}
y''=N(x, y, y') \label{eqna82}
\end{equation}
has a Lagrangian ${\cal L}(x, y, y')$ then (\ref{eqna82}) is equivalent to the Euler-Lagrange equation
\[\frac{d}{dx}\left(\frac{\partial {\cal L}}{\partial y'}\right)-\frac{\partial {\cal L}}{\partial y}=0\]
The determining equation for a Noether point symmetry
\begin{equation}
X=\xi(x,y)\frac{\partial}{\partial x}+\eta(x,y)\frac{\partial}{\partial y} \label{eqna84}
\end{equation}
corresponding to a Lagrangian ${\cal L}(x, y, y')$ of (\ref{eqna82}) is
\[X^{[1]}{\cal L}+\left(\frac{d\xi}{dx}\right){\cal L}=\frac{dF}{dx}\]
where 
\[ X^{[1]} = X + \left( \frac{d\eta}{dx} - y' \frac{d\xi}{dx} \right) \frac{\partial}{\partial y'} \] and
$F=F(x,y)$ is a gauge function.  It is also known that if there is a Noether symmetry corresponding to  a 
Lagrangian of an equation (\ref{eqna82}), then (\ref{eqna82}) can be reduced to quadratures.  This is a critical advantage 
of Noether point symmetries as emphasised by Wafo Soh and Mahomed \cite{Wafo1}.  The Noether first 
integral $I_N$ associated with the Noether point symmetry (\ref{eqna84}) is given by
\begin{equation}
I_N=\xi(x,y){\cal L}+\left(\eta(x,y)-y'\xi(x,y)\right){\cal L}_{y'}-F \label{eqna96}
\end{equation}
in terms of ${\cal L}$ and $F$.

For equation (\ref{eqn013}) a Lagrangian is
\begin{equation} {\cal L}=\frac{1}{2}y'^{2}+\frac{1}{3}f(x)y^{3}+\frac{1}{4}g(x)y^{4}\label{lag} \end{equation}
The Lagrangian ${\cal L}$ admits the Noether point symmetry
\begin{equation}
G=a\frac{\partial}{\partial x}+\left(by+c\right)\frac{\partial}{\partial y} \label{new}
\end{equation}
provided
\numparts
\begin{eqnarray}
b=\frac{1}{2}a' \label{eqna89a1}\\
g=g_{1}a^{-3} \label{eqna89a}\\
f=a^{-5/2}\left(f_{1}-3g_{1}\int ca^{-3/2}dx\right) \label{eqna89b}\\
a'''=4a^{-5/2}c\left(f_{1}-3g_{1}\int ca^{-3/2}dx\right) \label{eqna89c}\\
c=C_{0}+C_{1}x\label{eqna89d}\\
F=\frac{1}{4}a''y^{2}+c'y \label{eqna89e}
\end{eqnarray}
\endnumparts
where $a=a(x)$, $b=b(x)$, $c=c(x)$ and $f_{1}$ and $g_{1}$ are constants.  From (\ref{eqna89b}) and (\ref{eqna89c}) we have
\[f=\frac{a'''}{4c}\]
When we differentiate (\ref{eqna89c}) once with respect to $x$ we obtain
\begin{equation}
caa^{(iv)}+\frac{5}{2}ca'a'''-ac'a'''=-12g_{1}c^{3}a^{-3} \label{eqna92}
\end{equation}
Applying the transformation
\begin{eqnarray}
X&=&\frac{C_{1}}{C_{0}+C_{1}x}, \qquad A=\frac{aC_{1}^{2}}{\left(C_{0}+C_{1}x\right)^{2}} \label{eqna47}
\end{eqnarray}
reduces (\ref{eqna92}) to
\begin{equation}
AA^{(iv)}+\frac{5}{2}A'A'''=-12g_{1}C_{1}^{2}A^{-3} \label{eqna93}
\end{equation}
which is in the form of (\ref{eqna20}).  

From (\ref{eqna96}) we have that (\ref{eqn013}) admits the Noether first integral
\begin{eqnarray}
I_N=a\left[\frac{1}{2}y'^{2}+\frac{1}{3}y^{3}a^{-(5/2)}\left(f_{1}-3g_{1}\int ca^{-(3/2)}dx\right)+\frac{1}{4}g_{1}a^{-3}y^{4}\right] \nonumber\\
 +\left(\frac{1}{2}a'y+c-ay'\right)y'-\frac{1}{4}a''y^{2}-c'y \label{eqna90}
\end{eqnarray}
Substituting (\ref{eqna89c}) in (\ref{eqna90}) yields
\begin{equation}
I_N=cy'-\frac{1}{4}a''y^{2}+\frac{1}{2}a'yy'-\frac{1}{2}ay'^{2}+\frac{aa'''}{12c}y^{3}+\frac{1}{4}g_{1}a^{-2}y^{4}-c'y \label{eqna94}.
\end{equation}
On comparing our first integral (\ref{eqna18}) and the Noether first integral (\ref{eqna94}) we observe that, with $c=1/4$, $f_0=f_1$ and $g_0=g_1$,
\[\tau_{0}(t)=4I_N\]
As a result, our earlier {\it ad hoc} approach yielded a first integral which is a special case of 
that obtained via Noether's theorem. This further implies that (\ref{eqna18}) admits the Noether symmetry
\[Y= 4 a\frac{\partial}{\partial x}+\left(2a' y+1\right)\frac{\partial}{\partial y}\]
thus supporting the results of our approach, and the fact that (\ref{eqn013}) could be reduced to quadratures.

\section{Lie analysis}
As a final attempt at solving (\ref{eqn013}) we undertake a Lie symmetry analysis \cite{Olver}-\cite{Anco}.  
Lie symmetries are usually a larger set of symmetries for a problem as opposed to the set of 
Noether symmetries.  However, the disadvantage is that no simple formula exists to find first
integrals associated with Lie symmetries -- direct integration of (often difficult) equations is 
usually needed. However, as we show here, this approach allows for a larger class of solutions than the Noether approach.

It is a simple matter to verify that (\ref{new})
is a symmetry of (\ref{eqn013}) and the relationship among the functions $a(x), b(x), c(x), f(x)$ 
and $g(x)$ is given by the following system of ordinary differential equations
\numparts
\begin{eqnarray}
a''=2b' \label{eqna36a}\\
b''=2fc \label{eqna36b}\\
c''=0 \label{eqna36c}\\
a f'+(2a'+b)f=-3cg \label{eqna36d}\\
a g'+(2a'+2b)g=0 \label{eqna36e}
\end{eqnarray}
\endnumparts
From (\ref{eqna36a})
\[2b=a'+\alpha\]
where $\alpha$ is an arbitrary constant, from (\ref{eqna36a}) and (\ref{eqna36b})
\begin{equation}
f=\frac{a'''}{4c}, \label{eqna38}
\end{equation}
from (\ref{eqna36c}) we have (\ref{eqna89d}) again,
 and finally from (\ref{eqna36e})
\begin{equation}
g=g_{2}a^{-3}\exp\left(-\int\frac{\alpha dx}{a}\right) \label{eqna40}
\end{equation}
where $g_{2}$ is an arbitrary constant.  By integrating (\ref{eqna36d}) and using (\ref{eqna40}) we obtain an alternative form of (\ref{eqna38}), and it is
\begin{equation}
f=a^{-5/2}\exp\left(-\int\frac{\alpha dx}{2a}\right)\left[f_{2}-3g_{2}\int ca^{-(3/2)}\exp\left(\int\frac{\alpha dx}{a}\right)dx\right] \label{eqna42}
\end{equation}
where $f_{2}$ is an arbitrary constant of integration.  From (\ref{eqna36d}), (\ref{eqna38}) and (\ref{eqna40}) we observe that $a$ is a solution of the equation
\begin{equation}
caa^{(iv)}+\left[c\left(\frac{5a'}{2}+\frac{\alpha}{2}\right)-c'a\right]a'''=-12g_{2}c^{3}a^{-3}\exp\left(-\int\frac{\alpha dx}{a}\right) \label{eqna41}
\end{equation}
Alternatively (\ref{eqna41})  may be obtained by equating equations (\ref{eqna38}) and (\ref{eqna42}) and differentiating once with respect to $x$.

Observe the similarity between the results obtained here in the case of Lie analysis and those obtained 
in the previous section on Noether symmetries.  The main difference is the occurrence of the parameter 
$\alpha$ in the Lie analysis.  Thus, we have a more general result (as expected) in the Lie analysis as 
compared to the Noether analysis. In addition, this result is even more general than that obtained in 
Kweyama {\it et al} \cite{Kweyama} as our $c(x)$ is nonconstant.

The transformation which converts the symmetry (\ref{new}) to $\displaystyle \frac{\partial}{\partial X}$ 
makes (\ref{eqn013}) autonomous.  The simplest expression of this transformation is
\begin{eqnarray}
X&=&\int\frac{dx}{a}, \qquad Y=y \exp\left(-\int\frac{b dx}{a}\right)-\int\frac{c}{a}\exp\left(-\int\frac{b dx}{a}\right)dx \label{eqna52}
\end{eqnarray}
Using (\ref{eqna52}) equation  (\ref{eqn013}) becomes
\begin{equation}
Y''+\alpha Y'+\left(M+\frac{\alpha^{2}}{4}\right)Y=f_{2}Y^{2}+g_{2}Y^{3}+N \label{eqna53}
\end{equation}
where $f_{2}$ and $g_2$ are arbitrary constants introduced in (\ref{eqna40}) and (\ref{eqna42}) respectively.  
The quantities $M$ and $N$ are arbitrary constants that arise in  integrations of (\ref{eqna41}) and are given by
\begin{equation}
M=\frac{1}{2}aa''-\frac{1}{4}a'^{2}-2f_{2}I+3g_{2}I^{2} \label{eqna54}
\end{equation}
and
\begin{eqnarray}
N=-a^{-1/2}\exp\left(-\int\frac{\alpha dx}{2a}\right)\left(ac'-\frac{1}{2}a'c+\frac{1}{2}\alpha c\right)
\nonumber \\-\left(\frac{1}{2}aa''
-\frac{1}{4}a'^{2}
+\frac{\alpha^{2}}{4}\right)I
+f_{2}I^{2}-2g_{2}I^{3} \label{eqna55}
\end{eqnarray}
where
\begin{equation}
I=\int ca^{-(3/2)}\exp\left(-\int\frac{\alpha dx}{2a}\right)dx \label{eqna56}
\end{equation}
Note that in the neutral perfect fluid case we must have $g_2=N=0$ and (\ref{eqna53}) reduces to that 
of Maharaj {\it et al} \cite{Maharaj}. However, it is difficult to make direct comparison to the results therein 
as the full equations are not always given, and the equation referencing is not always clear.

To proceed further, we need to analyse equation (\ref{eqna41}).  We note that the form of (\ref{eqna41}) 
was obtained under the assumption that $c\neq0$.  We consider both nonzero $c$ and vanishing $c$  
in turn in our subsequent analysis.

\subsection{Case I: $c\neq0$}

When $C_{0}\neq 0$ and $C_{1}=0$, (\ref{eqna41}) may be written as
\begin{equation}
aa^{(iv)}+\frac{1}{2}\left(5a'+\alpha\right)a'''=-12g_{2}a^{-3}\exp\left(-\int\frac{\alpha dx}{a}\right) \label{eqna43}
\end{equation}
Rescaling $a$ and $x$ in (\ref{eqna43}) yields
\begin{equation}
aa^{(iv)}+\frac{1}{2}\left(5a'+\alpha\right)a'''=a^{-3}\exp\left(-\int\frac{\alpha dx}{a}\right) \label{eqna44}
\end{equation}
when $\alpha \neq 0$ and
\begin{equation}
aa^{(iv)}+\frac{5}{2}a'a'''=a^{-3} \label{eqna45}
\end{equation}
when $\alpha =0$.

When $C_{0}\neq 0$ and $C_{1}\neq 0$, we apply transformation (\ref{eqna47}) to (\ref{eqna41}) and 
we obtain (again with the rescaling of $A$ and $X$)
\begin{equation}
AA^{(iv)}+\frac{1}{2}\left(5A'-\alpha\right)A'''=A^{-3}\exp\left(\int\frac{\alpha dX}{A}\right) \label{eqna48}
\end{equation}
when $\alpha \neq 0$ and
\[AA^{(iv)}+\frac{5}{2}A'A'''=A^{-3}\]
when $\alpha = 0$.  Changing the sign of $\alpha$ in (\ref{eqna48}) brings it to the form of (\ref{eqna44}), 
and so the critical equations are (\ref{eqna44}) and (\ref{eqna45}).

\subsubsection{Case Ia: $\alpha=0$.}
\label{azero}
If $\alpha =0$ then (\ref{eqna53}) becomes
\begin{equation}
Y''+MY=f_{2}Y^{2}+g_{2}Y^{3}+N \label{eqna57}
\end{equation}
The solution of (\ref{eqna57}) is expressed as the quadrature
\begin{equation}
X-X_{0}=\int \frac{dY}{\sqrt{\frac{1}{2}g_{2}Y^{4}+\frac{2}{3}f_{2}Y^{3}-MY^{2}+2NY+L}} \label{eqna58}
\end{equation}
where $L$ is an arbitrary constant introduced in the first integration of (\ref{eqna57}). A full discussion of the 
evaluation of this quadrature can be found in Kweyama {\it et al} \cite{Kweyama}. If we invoke the 
same transformations in obtaining (\ref{eqna57}) to the first integral (\ref{eqna94}) we obtain
\[-L=\frac{1}{4}g_{1}Y^{4}+\frac{1}{3}f_{1}Y^{3}-\frac{1}{2}MY^{2}+NY-\frac{1}{2}Y'^{2}\]
which is the intermediate integration between (\ref{eqna57}) and (\ref{eqna58}).  This
again indicates that the Noether results are a subset of the Lie results.

We now have to determine $a(x)$.  We make the observation that if $\alpha=0$ and $g_{0}=16g_{2}$ and 
$g_{1}=g_{2}/C_{1}^{2}$, (\ref{eqna20}) and (\ref{eqna93}) become (\ref{eqna43}) which reduces to (\ref{eqna45}).  
We therefore consider (\ref{eqna45}) for further analysis and reduction to quadrature.  In carrying out the 
Lie analysis of (\ref{eqna45}), using PROGRAM LIE \cite{Head}, we find that it has two Lie point symmetries, namely
\begin{eqnarray*}
G_{1}&=&\frac{\partial}{\partial x} \\
G_{2}&=&x\frac{\partial}{\partial x}+\frac{4}{5}a\frac{\partial}{\partial a}
\end{eqnarray*}
Usually, when an $n$th order equation admits an $m<n$ dimensional Lie algebra of symmetries, there is little 
hope for the solution of the equation via those symmetries.  However, in this case we are able to reduce 
the equation due to the presence of hidden symmetries \cite{asg}.

The symmetry $G_{1}$ determines   the variables for reduction
\begin{eqnarray*}
u&=&a, \qquad v=a'
\end{eqnarray*}
and the reduced equation is
\[u^{4}v^{3}v'''+4u^{4}v^{2}v'v''+\frac{5}{2}u^{3}v^{3}v''+u^{4}vv'^{3}+\frac{5}{2}u^{3}v^{2}v'^{2}-1=0 \]
This equation admits the following two symmetries
\begin{eqnarray*}
U_{1}&=&u\frac{\partial}{\partial u}-\frac{1}{4}v\frac{\partial}{\partial v}\\
U_{2}&=&2u^{2}\frac{\partial}{\partial u}+uv\frac{\partial}{\partial v}
\end{eqnarray*}
The variables for reduction via $U_{2}$ are
\begin{eqnarray*}
r&=&u^{-\frac{1}{2}}v \qquad s=u^{\frac{3}{2}}v'-\frac{1}{2}u^{\frac{1}{2}}v
\end{eqnarray*}
and the reduced equation is
\begin{equation}
r^{3}s^{2}s''+r^{3}ss'^{2}+4r^{2}s^{2}s'+rs^{3}-1=0 \label{eqnaa28}
\end{equation}
The Lie symmetry analysis of (\ref{eqnaa28}) yields the following two symmetries
\begin{eqnarray*}
X_{1}&=&r\frac{\partial}{\partial r}-\frac{s}{3}\frac{\partial}{\partial s}\\
X_{2}&=&\frac{\partial}{\partial r}-\frac{s}{r}\frac{\partial}{\partial s}
\end{eqnarray*}
The reduction variables generated by  $X_{2}$ are
\begin{eqnarray*}
p&=&rs \qquad q=rs'+s \label{eqnaa31}
\end{eqnarray*}
and the reduced equation is
\[p^{2}q q'+pq^{2}-1=0 \]
with solution
\[q^{2}=\frac{2}{p}+\frac{q_0}{p^{2}} \]
where $q_0$ is a constant. We can now invert these transformations to find the solution of (\ref{eqna45}).  
Alternatively, we can integrate (\ref{eqna45}) directly and write down the solution as
\begin{eqnarray*}
u_{x}&=&{a}^{-3/2}=[G'(u)]^{-1}\\
x-x_{0}&=&G(u)
\end{eqnarray*}
where we have set
\[G(u)=\int\frac{du}{\left({\cal K}_{4}+{\cal K}_{3}u+{\cal K}_{2}u^{2}-(1/6){\cal K}_{1}u^{3}+(1/32){\cal K}_{0}u^{4}\right)^{3/2}}\]
and the ${\cal K}_i, i=0,\ldots,3$ are constants of integration related to $M, N, f_2$ and $g_2$ and ${\cal K}_4$ is  arbitrary.
As pointed out earlier, this result was obtained in Kweyama {\it et al} \cite{Kweyama}, but for constant $c(x)$.  In the case of
 nonconstant $c(x)$, the solution is the same, except that we replace $a$ and $x$ in this solution with $A$ and $X$ respectively.  
 To obtain the solution to (\ref{eqna41}) (with $\alpha=0$) we need to apply the inverse of (\ref{eqna47}).

\subsubsection{Case Ib: $\alpha\neq0$.}

When $\alpha\neq 0$, we cannot directly reduce (\ref{eqna53}) to quadratures.  We need to investigate the constraints 
under which it possesses a second point symmetry.   We find that if $f_{2}\neq0$, $g_{2}\neq0$ then (\ref{eqna53})
 has the following two symmetries
\numparts
\begin{eqnarray}
G_{1}&=&\frac{\partial}{\partial X} \label{eqna59a}\\
G_{2}&=&e^{(\alpha/3)X}\frac{\partial}{\partial X}-e^{(\alpha/3)X}\left(\frac{\alpha}{3}Y+\frac{\alpha f_{2}}{9g_{2}}\right)\frac{\partial}{\partial Y} \label{eqna59b}
\end{eqnarray}
\endnumparts
provided the following conditions are satisfied
\begin{equation}
M=-\frac{f_{2}^{2}}{3g_{2}}-\frac{\alpha^{2}}{36}, \qquad N=\frac{f_{2}^{3}}{27g_{2}^{2}}-\frac{2\alpha^{2}f_{2}}{27g_{2}} \label{mnconds}
\end{equation}
Utilising (\ref{eqna59b}) we obtain the transformation
\begin{eqnarray*}
\cal X&=&-\frac{3}{\alpha}e^{-(\alpha/3)X},  \qquad {\cal Y}=e^{(\alpha/3)X}\left(Y+\frac{f_{2}}{3g_{2}}\right)
\end{eqnarray*}
and equation (\ref{eqna53}) becomes
\[{\cal Y}''=g_{2}{\cal Y}^{3} \]
with  solution
\[{\cal X}-{\cal X}_{0}=\int \frac{d{\cal Y}}{\sqrt{\frac{g_{2}}{2}{\cal Y}^{4}+{\cal C}}} \]
The values in (\ref{mnconds}) look rather peculiar.  However, it is interesting to note that these values correspond directly to a simplification of the eigenvalue problem associated with a dynamical systems analysis of (\ref{eqna53})!

When $c\neq0$, $g_{2}=0$, $f_{2}\neq0$ we find that $g\equiv0$ (and so we are in the neutral perfect fluid realm). 
 Now  (\ref{eqna53}) has the following two symmetries
\numparts
\begin{eqnarray}
Y_{1}&=&\frac{\partial}{\partial X} \label{eqna67a}\\
Y_{2}&=&e^{(\alpha/5)X}\frac{\partial}{\partial X}+\left(\frac{\alpha^{3}}{500f_{2}}+\frac{\alpha M}{5f_{2}}-\frac{2\alpha}{5}Y\right)e^{(\alpha/5)X}\frac{\partial}{\partial Y} \label{eqna67b}
\end{eqnarray}
\endnumparts
provided the following condition is satisfied
\begin{eqnarray}
N&=& \frac{M^{2}}{4f_{2}}+\frac{\alpha^{2}M}{8f_{2}}+\frac{49\alpha^{4}}{40 000f_{2}}\nonumber\\
&=&\frac{1}{4f_{2}}\left(M+\frac{\alpha^{2}}{4}\right)^{2}-\frac{36\alpha^{4}}{2500f_{2}} \label{eqna67}
\end{eqnarray}
This condition is equivalent to the one obtained by Mellin {\it et al} \cite{Mellin} for the case where $n=2$ in their 
analysis of the generalised Emden-Fowler equation.  We use (\ref{eqna67b}) to obtain the following transformation
\begin{eqnarray*}
{\cal X}&=&-\frac{5}{\alpha}e^{(-\alpha/5)X},  \qquad {\cal Y}=\left(Y-\frac{M}{2f_{2}}-\frac{\alpha^{2}}{200f_{2}}\right)e^{(2\alpha/5)X}
\end{eqnarray*}
which, together with (\ref{eqna67}) reduces equation (\ref{eqna53}) to
\[{\cal Y}''=f_{2}{\cal Y}^{2} \]
with solution
\[{\cal X}-{\cal X}_{0}=\int \frac{d{\cal Y}}{\sqrt{\frac{2f_{2}}{3}{\cal Y}^{3}+{\cal C}_{1}}}\]
We take this opportunity to make two minor corrections to the work of
 Mellin {\it et al} \cite{Mellin}: While the expression for $N$ given by their equation (7.11) is correct, it is obtained by  
 multiplying  their equation (7.9) by $a\int da^{-3/2}\exp \left[\frac{1}{2}\int\left(p-2C_{0}/a\right)dx\right]dx$ and then 
 integrating, not multiplying by $a\int da^{-3/2} dx$ as indicated in their paper.  Also, the coefficient of $C_{0}$
  should be $2$ in their equation (7.8).

\subsection{Case II: $c=0$}
\label{cons}

From (\ref{eqna36a}), (\ref{eqna36b}), (\ref{eqna36d}) and (\ref{eqna36e}) we have
\numparts
\begin{eqnarray}
b&=&\frac{1}{2}\left(a'+\alpha\right) \label{eqn014a}\\
a&=&a_{0}+a_{1}x+a_{2}x^{2} \label{eqn014b}\\
f&=&f_{2}a^{-5/2}\exp\left(-\int \frac{\alpha dx}{2a}\right) \label{eqn014c}\\
g&=&g_{2}a^{-3}\exp \left(-\int \frac{\alpha dx}{a}\right) \label{eqn014d}
\end{eqnarray}
\endnumparts
The symmetry (\ref{new}) now takes the form
\[G=a\frac{\partial}{\partial x}+\frac{1}{2}\left(a'+\alpha\right)y\frac{\partial}{\partial y} \]
Using the transformation
\begin{eqnarray*}
X&=&\int \frac{dx}{a}, \qquad Y=ya^{-1/2}\exp\left(-\int \frac{\alpha dx}{2a}\right)
\end{eqnarray*}
equation (\ref{eqn013}) is transformed into the autonomous equation
\begin{equation}
Y''+ \alpha Y'+\beta Y=f_{2}Y^{2}+g_{2}Y^{3} \label{eqn017}
\end{equation}
where
\begin{eqnarray*}
\beta&=&\frac{1}{4}\left(\alpha^{2}-\Delta \right), \qquad \Delta=a_{1}^{2}-4a_{0}a_{2}
\end{eqnarray*}
In carrying out the standard Lie point symmetry analysis on (\ref{eqn017}) we have the following cases:

\subsubsection{Case IIa.}
If $f_{2}\neq0$, $g_{2}\neq0$, (\ref{eqn017}) has the following two symmetries
\numparts
\begin{eqnarray}
G_{1}&=&\frac{\partial}{\partial X} \label{eqn019} \label{eqn019a}\\
G_{2}&=&e^{(\alpha/3)X}\frac{\partial}{\partial X}-e^{(\alpha/3)X}\left(\frac{\alpha}{3}Y+\frac{2\alpha^{3}}{9f_{2}}\right)\frac{\partial}{\partial Y} \label{eqn019b}
\end{eqnarray}
\endnumparts
provided the following conditions apply
\begin{eqnarray}
\beta&=&-\frac{4\alpha^{2}}{9}, \qquad g_{2}=\frac{f_{2}^{2}}{2\alpha^{2}} \label{eqn020}
\end{eqnarray}
We use (\ref{eqn019b}) to obtain the following transformation
\begin{eqnarray}
{\cal X}&=&-\frac{3}{\alpha}e^{-(\alpha/3)X}, \qquad {\cal Y}=e^{-(\alpha/3)X}\left(Y+\frac{2\alpha^{2}}{3f_{2}}\right) \label{eqn021}
\end{eqnarray}
Using (\ref{eqn020}) and (\ref{eqn021}) the equation (\ref{eqn017}) becomes
\begin{equation}
{\cal Y''}=\frac{f_{2}^{2}}{2\alpha^{2}}{\cal Y}^{3} \label{eqn022}
\end{equation}
and the solution of (\ref{eqn022}) is
\[{\cal X}-{\cal X}_{0}=\int\frac{d{\cal Y}}{\sqrt{\frac{f_{2}^{2}}{4\alpha^{2}}{\cal Y}^{4}+{\cal C}}} \]

\subsubsection{Case IIb.}

If $f_{2}=0$, $g_{2}\neq0$ (which implies that $f=0$), then (\ref{eqn017}) has the following two symmetries
\numparts
\begin{eqnarray}
G_{1}&=&\frac{\partial}{\partial X} \label{eqn024}\\
G_{2}&=&e^{(\alpha/3)X}\frac{\partial}{\partial X}-\frac{\alpha}{3}e^{(\alpha/3)X}Y\frac{\partial}{\partial Y} \label{eqn025}
\end{eqnarray}
\endnumparts
subject to the following condition
\begin{equation}
\beta=\frac{2\alpha^{2}}{9} \label{eqn026}
\end{equation}
Using (\ref{eqn025}) we obtain the following transformation
\begin{eqnarray}
{\cal X}=-\frac{3}{\alpha}e^{-(\alpha/3)X}, \qquad {\cal Y}=e^{(\alpha/3)X}Y \label{eqn027}
\end{eqnarray}
Using (\ref{eqn026}) and (\ref{eqn027}) equation (\ref{eqn017}) is transformed to
\begin{equation}
{\cal Y}''=g_{2}{\cal Y}^{3} \label{eqn028}
\end{equation}
and the solution of (\ref{eqn028}) is
\[{\cal X}-{\cal X}_{0}=\int \frac{d{\cal Y}}{\sqrt{\frac{g_{2}}{2}{\cal Y}^{4}+{\cal C}}} \]
This is an intrinsically charged result - there is no uncharged analogue. \label{charge}

\subsubsection{Case IIc.}

If $f_{2}\neq0$, $g_{2}=0$ (which implies $g=0$), then (\ref{eqn017}) has two symmetries provided
\begin{equation}
\beta=\pm\frac{6\alpha^{2}}{25} \label{eqn035}
\end{equation}
and can be transformed to
\[{\cal Y}''=f_{2}{\cal Y}^{2} \]
which has the solution
\[{\cal X}-{\cal X}_{0}=\int \frac{d{\cal Y}}{\sqrt{\frac{2f_{2}}{3}{\cal Y}^{3}+{\cal C}}}\]
This result was previously obtained in the neutral case by Maharaj {\it et al} \cite{Maharaj}.
A {\it consequence} of (\ref{eqn035}) is that $\Delta>0$ (Note that this is not imposed on (\ref{eqn017}) as 
was done in Maharaj {\it et al} \cite{Maharaj}.) and hence $a$ in (\ref{eqn014b}) has real roots.

\section{Discussion}
We have undertaken a comprehensive analysis of (\ref{eqn013}) in order to determine 
which forms of the functions $f$ and $g$ would lead to first integrals and/or solutions of the
 equation.  We reviewed our previous {\it ad hoc} approach, and showed that those results 
 were contained in the results obtained via Noether's theorem.  This occurred when the 
 function $c$ obtained in the Noether analysis was set to $\frac14$. These latter results were 
 then shown to be further contained with those obtained via the Lie analysis. This occurs 
 when $\alpha =0$.  In this case (\ref{eqna40}) and (\ref{eqna42}) take the forms of 
 (\ref{eqna89a}) and (\ref{eqna89b}) respectively, and so (\ref{eqna89a}) and (\ref{eqna89b}) 
 are special cases of (\ref{eqna40}) and (\ref{eqna42}) respectively.

Usefully, the first two approaches, namely the \textit{ad hoc} approach and the 
Noether symmetry analysis, yielded first integrals directly.  The final (Lie) approach needed
 two stages of analysis in order to reduce the equation to quadratures.

We were able to completely analyse (\ref{eqn013}) for Lie point symmetries in an exhaustive 
analysis.  All possible cases for the functions $f$ and $g$ were analysed.  We showed that 
these results reduced to the neutral case results ($g=0$) of Maharaj {\it et al} \cite{Maharaj} 
in most cases.  However, we were also able to find an inherently charged case in \S \ref{charge} 
that has no uncharged analogue.

While a complete analysis was produced for the case of $\alpha =0$, only a partial analysis could 
be performed when $\alpha\neq0$. Nonetheless, we were still able to provide conditions 
under which (\ref{eqn013}) could be reduced to quadratures. It still remains to solve (\ref{eqna44}).  
The main difficulty is that this equation in an integro-differential equation and such equations 
are notoriously difficult to solve. However, we can still reduce (\ref{eqna44}) to a first order 
equation by eliminating $a''$ from (\ref{eqna54}) and (\ref{eqna55}). Further work in this direction is ongoing.

Note that, in the case that $c=0$, we were able to find constraints under which we could reduce 
(\ref{eqn013}) to quadratures with $f$ and $g$ given explicitly and $\alpha\neq0$.  It is interesting 
to observe that  the constraints we found in all subcases of \S \ref{cons} forced the quadratic $a$ in (\ref{eqn014b})  to have real roots.

To complete our analysis, we take the symmetries calculated in \S \ref{azero} and investigate the possibility of 
group invariant solutions \cite{Bluman, Anco} of (\ref{eqna45}).  The only two results of significance are that
\begin{equation}
x-x_{0}=\pm\int\frac{da}{\left[-\left(\frac{32}{3}\right)^{1/3}a^{-(1/3)}+K_{1}a^{2/3}\right]^{3/4}}\label{eqnn1}
\end{equation}
and
\begin{equation}
a=-\left(\frac{625}{24}\right)^{1/5}x^{4/5}\label{eqnn2}
\end{equation}
are invariant solutions of (\ref{eqna45}).  The first solution can only be given implicitly and so is not of much practical use.  
However, if we let $c=\frac{1}{4}$ in (\ref{eqna38}), $\alpha=0$ in (\ref{eqna40}) and $K_{1}=0$ in (\ref{eqnn1}), then
  we obtain explicit forms of $f$ and $g$.  Also if we substitute (\ref{eqnn2}) into (\ref{eqna38})  and (\ref{eqna40}) we can
  explicitly determine $f$ and $g$. In both cases we obtain the forms
\begin{equation}
\label{eqnn3}
f(x) \propto \left(x-x_{0}\right)^{-(11/5)}, \qquad
g(x) \propto \left(x-x_{0}\right)^{-(12/5)}
\end{equation}
These forms for $f$ and $g$ were earlier obtained by Kweyama \textit{et al} \cite{Kweyama} using an \textit{ad hoc} 
approach which yielded a new charged first integral to the Einstein-Maxwell field equations.  In this case, (\ref{eqn013}) admits two Lie point symmetries and can be reduced to quadratures.

  We can also reuse (\ref{eqnn2}) by invoking (\ref{eqna47}) to obtain
\begin{equation}
a = \kappa \left(\frac{c'}{c}\right)^{-(6/5)}
\end{equation}
which is a solution to (\ref{eqna41}) provided
\begin{equation}
g_2= -\frac{24 \kappa^5}{625 C_1^2}.
\end{equation}
This corresponds to the choice
\begin{equation}
\label{eqnn3}
f(x) \propto \left(x-x_{0}\right)^{-(14/5)}, \qquad
g(x) \propto \left(x-x_{0}\right)^{-(18/5)}
\end{equation}
for which (\ref{eqn013}) again has two Lie point symmetries and can be reduced to quadratures.

In summary, we have given a complete Noether and Lie point symmetry analysis of  (\ref{eqn013}).   For the Lagrangian (\ref{lag}) we were able to provide the most general Noether point symmetry, the most general first integral associated with this symmetry and indicated that this integral was equivalent to that found via an {\it ad hoc} approach, {\it i.e} (\ref{eqna18}).  Finally we determined the most general Lie point symmetry admitted by (\ref{eqn013}) and gave conditions under which the equation could be reduced  either to a first order equation, or to quadratures.

~\\
{\bf Acknowledgements}\\
MCK and KSG thank the National Research Foundation and the
University of KwaZulu-Natal for financial support. SDM acknowledges
that this work is based upon research supported by the South African
Research Chair Initiative of the Department of Science and
Technology and the National Research Foundation. 
~\\

\end{document}